\documentclass[runningheads]{llncs}

\usepackage{multirow}
\usepackage{algorithm}
\usepackage{algpseudocode}
\usepackage{graphicx}
\usepackage{xcolor}
\usepackage{soul}
\usepackage{subfig}
\usepackage{amsmath}
\usepackage{amsfonts}
\usepackage{caption}
\usepackage[para]{footmisc}
\usepackage{booktabs}
\usepackage[noadjust]{cite}
\usepackage{xcolor,colortbl}
\usepackage{hyperref}
\hypersetup{colorlinks = true, citecolor = blue }
\usepackage{cleveref}
\usepackage{caption}
\usepackage{bbding}
\usepackage[misc]{ifsym}

\usepackage{svg}
\newcommand{\orcid}[1]{\href{https://orcid.org/#1}{\includesvg[width=10pt]{orcid}}}

\newcommand\blfootnote[1]{%
  \begingroup
  \renewcommand\thefootnote{}\footnote{#1}%
  \addtocounter{footnote}{-1}%
  \endgroup
}

\usepackage{orcidlink}

\begin{document}
\title{Explaining AI Decisions: Towards Achieving Human-Centered Explainability in Smart Home Environments}

\titlerunning{Towards Achieving Human-Centered Explainability in Smart Home}

\author{Md Shajalal (\Letter) \inst{1,2}  \orcidlink{0000-0002-9011-708X}\and
Alexander Boden\inst{1,3}\orcidlink{0000-0002-6470-1151}\and
Gunnar Stevens\inst{2,3} \orcidlink{0000-0002-7785-5061} \and
Delong Du\inst{2}\orcidlink{0000-0001-8916-3524} \and
Dean-Robin Kern\inst{2}\orcidlink{0000-0001-9810-3013}
\institute{
Fraunhofer Institute for Applied Information Technology FIT, Sankt Augustin, Germany \\
\email{md.shajalal@fit.fraunhofer.de}
\and 
University of Siegen, Siegen, Germany
\and
Bonn-Rhein-Sieg University of Applied Sciences, Sankt Augustin, Germany
}
}
\authorrunning{Md Shajalal et al.}

\maketitle              
\begin{abstract}
Smart home systems are gaining popularity as homeowners strive to enhance their living and working environments while minimizing energy consumption. However, the adoption of artificial intelligence (AI)-enabled decision-making models in smart home systems faces challenges due to the complexity and black-box nature of these systems, leading to concerns about explainability, trust, transparency, accountability, and fairness. The emerging field of explainable artificial intelligence (XAI) addresses these issues by providing explanations for the models' decisions and actions. While state-of-the-art XAI methods are beneficial for AI developers and practitioners, they may not be easily understood by general users, particularly household members. This paper advocates for human-centered XAI methods, emphasizing the importance of delivering readily comprehensible explanations to enhance user satisfaction and drive the adoption of smart home systems. We review state-of-the-art XAI methods and prior studies focusing on human-centered explanations for general users in the context of smart home applications. Through experiments on two smart home application scenarios, we demonstrate that explanations generated by prominent XAI techniques might not be effective in helping users understand and make decisions. We thus argue for the necessity of a human-centric approach in representing explanations in smart home systems and highlight relevant human-computer interaction (HCI) methodologies, including user studies, prototyping, technology probes analysis, and heuristic evaluation, that can be employed to generate and present human-centered explanations to users.                  
\blfootnote{This is the pre-print version of our accepted paper at the 2nd World Conference on eXplainable Artificial Intelligence (xAI2024), which will be held in Valletta, Malta in 17-19 July, 2024}                                       
 
\end{abstract}
\section{Introduction}
Due to advancements in sensor technology and machine learning (ML) over the past decades, smart home applications can now provide residents with the ability to monitor and control connected appliances via sensors~\cite{shajalal2022towards}. These applications can even make decisions automatically using ML-driven techniques rather than relying on simple timetable logic. In the smart home energy domain, one notable energy-aware smart home application might be appliance-level energy-demand forecasting to make users more aware and help them optimize their energy consumption practices~\cite{shajalal2022towards,kim2020electric}. Adjusting the heating system to provide a comfortable and healthy household and work environment based on predicting individuals' thermal comfort preferences can be another fascinating energy-related smart home application~\cite{alan2016too,shajalal2024improved,shajalal2022focus}. Other applications also often utilize complex ML models to make decisions, such as human activity recognition within the home, identification of energy-intensive activities for different household tasks~\cite{stankovic2016measuring}, fall detection and health monitoring~\cite{mshali2018survey}, and energy optimization~\cite{kim2021explainable,shajalal2022towards}.

AI-based applications in smart home systems are becoming increasingly popular as homeowners aim to enhance their living environment while reducing energy usage. Previous studies~\cite{kim2019electric, kim2020electric, kim2021explainable, vanting2021scoping,shajalal2022towards} have modeled energy demand forecasting in smart homes using AI techniques, including Deep Neural Networks~(DNNs), Convolutional Neural Networks~(CNNs), Auto-Encoders~(AE), and Long-Short-Term Memory~(LSTM). Classical Machine Learning~(ML)-based predictive models have also garnered attention for predicting personal thermal comfort in indoor environments, thereby enhancing resident comfort~\cite{shajalal2024improved,abdelrahman2022personal, chennapragada2022time, eslamirad2020thermal, gao2021transfer, somu2021hybrid, sakkas2023explainable, quintana2020balancing}. The complexity and opacity of these ML models often hinder their adoption in real-world scenarios due to difficulties in aiding users' decision-making. Since ML models can be very complex, involving thousands to millions of model parameters (i.e., deep learning models), they are often referred to as \emph{black-boxes}. Decisions from black-box models can be unintelligible and may surprise users with unexpected predictions. In such cases, users require explanations to comprehend the predictions. Recently, there has been significant interest in elucidating the decisions of ML models across various fields, including language processing~\cite{shajalal2023unveiling}, financial analytics ~\cite{shajalal2022explainable}, e-commerce~\cite{shajalal2022explainable}, medicine and health~\cite{karim2023interpreting}, bioinformatics~\cite{karim2023explainable}, and smart home applications~\cite{kim2021explainable,karim2023explainable,shajalal2022towards}. This effort to clarify ML models' decisions is referred to as eXplainable Artificial Intelligence~(XAI).

XAI aims to develop AI systems that can provide clear explanations about the decision-making processes and the predicted decisions~\cite{arrieta2020explainable}. The ''black-box'' issue can lead to users' mistrust and confusion about these technologies. To improve users' trust and understanding, \emph{Human-centered} explanations\footnote{Throughout the paper, ``Human-centered XAI'' and ``user-centered XAI'' are used interchangeably} can be a game-changer by providing clear and effective explanations to users, enabling them to troubleshoot issues and customize devices to suit their needs~\cite{ehsan2022human}. However, many XAI methods have been introduced and developed to explain the models' decision-making procedures and the reasons behind specific predictions. Most of them are proposed to debug and improve models' performance~\cite{shajalal2022towards, ehsan2021operationalizing, kabir2021explainable}. In the context of smart home application scenarios, users in households are generally laypeople and may not have sufficient knowledge to understand technical explanations (i.e., even AI developers might struggle to understand explanations). Therefore, this paper advocates the need for easily understandable explanations for general users to make sense of predictions, which we refer to as ``Human-centered'' explanations.

While some studies have attempted to make models interpretable in the context of smart homes, most are focused on explaining predictions to improve performance and debug models~\cite{mucha2020towards, grimaldo2020combining, kim2020electric}. As noted earlier, smart home users often lack the technical expertise necessary to understand many of the suggested explanations~\cite{riboni2021keynote, rai2020explainable, kabir2021explainable}. Additionally, research has shown that end-users have diverse perspectives when trying to make sense of smart home systems~\cite{castelli2017happened}, and these perspectives can evolve over time. Various studies have demonstrated that current XAI techniques fail to produce human-centered explanations that assist general users in understanding the decision-making process and the reasons behind specific predictions~\cite{rong2022towards, bell2022s}. Given these findings, generating human-centered explanations for complex smart home applications is more challenging than might be initially assumed, especially considering the diverse backgrounds of general users. 

Smart home systems encompass various energy consumption-related subtasks, including energy demand forecasting~\cite{shajalal2022towards,kim2021explainable}, appliance-level consumption predictions~\cite{kim2019electric}, energy intensity identification for different household activities~\cite{stankovic2016measuring}, and thermal comfort prediction~\cite{shajalal2022focus,shajalal2024improved} for efficient heating systems. These systems are more complex than classical classification or regression tasks, and their collective outcomes contribute to the overall functionality of smart homes. However, the complexity of these systems can lead to decisions that are difficult for general users to understand, potentially hindering their adoption in real-world settings. This paper emphasizes the importance of human-centered explanations in smart home applications by reviewing the progress of state-of-the-art technical and human-centered XAI studies. We focus on two energy consumption-related application scenarios within smart home settings: energy demand forecasting and the prediction of personal thermal comfort preferences for smart heating systems. We conduct experiments by applying deep neural network-based energy forecasting and ML models on benchmark datasets to model thermal comfort preferences. To uncover the complexity of the predictions, we apply current prominent XAI methods to identify facts by presenting the explanations in various forms.

We present explanations generated by multiple XAI methods and analyze their understandability. We then identify the associated challenges that must be considered when generating human-centered explanations. The need for user-friendly explanations in these contexts illustrates the challenges in understanding complex decisions. Finally, we highlight several HCI methodologies that could be beneficial in achieving human-centered XAI in smart home applications. The contributions of this paper are threefold: 
\begin{itemize}
    \item We present state-of-the-art XAI methods, discuss progress towards human-centered XAI and highlight research gaps that hinder their immediate application in smart home systems.
    \item Through careful analysis of experimental results based on prominent explainability methods in two sub-tasks, we argue for the need for human-centered explainability to understand decisions from complex AI-enabled smart home applications.
    \item We also emphasize several human-computer interaction (HCI) methodologies, including user studies, prototyping, technology probes analysis, and heuristic evaluation, to achieve human-centered XAI-enabled smart home applications.
\end{itemize}

The rest of the paper is organized as follows: In Section~\ref{HCXAI-SOTA}, we present state-of-the-art technical XAI methods and human-centered XAI studies. In Section~\ref{application-scenario}, we present two smart home application scenarios that illustrate the need for human-centered explainability in smart homes. We conducted experiments applying prominent ML techniques and XAI methods to predict and explain the models' decisions in Section~\ref{experiments_xai}. We also demonstrate why the provided explanations are insufficient for users to understand them. In Section~\ref{challenge_direction}, we highlight multiple HCI methodologies that demonstrate how to elicit requirements and design human-centered explanations. Finally, Section~\ref{conclusion} concludes the paper by outlining future directions.

\section{Human-Centered XAI and Current Progress} \label{HCXAI-SOTA}

The broad goal of XAI is to enable general users to understand the working principles of AI models and their decisions through explanations. Several terms, such as ``\emph{interpretable AI}'' and ``\emph{transparent AI}'', are used interchangeably to describe the exact purpose of XAI~\cite{rong2022towards}. The major objectives are similar: to make AI models and their decision-making processes understandable to general users through explanations. However, the number of methods focusing on a user-centered perspective is significantly lower than methods prioritizing improving model performance with technical explanations. As a result, the broad goal of XAI has not yet been fully achieved, which could potentially hinder the adoption of AI models in real-world applications.

This section provides a concise overview of the current cutting-edge approaches in explaining opaque decisions made by machine learning and deep learning-based predictive technologies, focusing on both technical and human-centered XAI. The primary aim of XAI advancements is to clarify the overall priorities and predictions of models for developers, facilitating the debugging process and enhancing model performance. While the field of technical XAI offers a wide range of techniques aimed at model improvement, it is important to note that the consideration of human-centered perspectives is relatively limited~\cite{rong2022towards}. To present a comprehensive understanding of the literature, this section focuses on three key issues: i) technical XAI, which provides a background of XAI; ii) human-centered XAI; iii) research gaps in the development of human-centered applications for smart home technology.

\subsection{Technical XAI}

The methods used in explainable artificial intelligence can be broadly classified into two categories: global and local explanation methods~\cite{molnar2020interpretable}. Global explainability methods aim to identify the overall priorities of a predictive model and provide a summary of the decision-making process. In contrast, local explainability methods focus on understanding why a specific predictive decision was made, shedding light on the insights associated with that decision. Additionally, explainable AI approaches can be categorized as either model-agnostic or model-specific. Model-agnostic approaches can be applied to any predictive model to explain its predictions, while model-specific explainable techniques are designed for particular predictive algorithms~\cite{shajalal2022explainable}.

One prominent and widely used XAI method is SHapley Additive Explanation (SHAP)~\cite{lundberg2017unified}, which generally explains the global priorities of models and highlights the most and least significant features according to their contribution. Following the game theory concept, SHAP computes each feature's weight that contributes to a specific decision. Moreover, SHAP can also provide local explanations for specific decisions. To explain deep neural network-based predictive models, well-known methods such as Grad-CAM~\cite{selvaraju2017grad}, based on gradient localization, and Layer-wise Relevance Propagation (LRP)~\cite{montavon2019layer}, which redistributes the output weight using backward propagation, can be used. These methods rely on saliency maps to explain decisions and can be applied to image and text-based applications. To uncover a DNN model's decisions, DeepLIFT (Deep Learning Important FeaTures)~\cite{shrikumar2017learning} has been introduced to identify important features for specific predictions of a model. Lakkaraju et al.~\cite{lakkaraju2019faithful} proposed an XAI technique for AI experts to make the model and its behavior understandable by using subspace explanations. To provide explanations for experts, Schetinin et al.~\cite{schetinin2007confident} introduced probabilistic interpretation for Bayesian decision tree models.

Another way to explain models' predictions to users is through example-based explanations. With these types of explanations, XAI approaches provide similar instances that match the corresponding samples with the same decision. Several example-based interpretability models that consider similar prototypes and criticisms have been introduced to help users understand why a certain decision was made~\cite{kim2016examples,gurumoorthy2019efficient}. To explain predictions for any complex deep learning model, Local Interpretable Model-Agnostic Explanations (LIME)~\cite{ribeiro2016should} can create a \emph{surrogate model} that mimics the performance of the complex model. The \emph{surrogate model} is explainable, and its performance is quite similar to that of the original model. With LIME, any particular prediction can be explained for tabular and textual data by highlighting positive and negative features corresponding to the predicted decision.

To comprehend specific events, humans sometimes look for explanations that involve significant changes in the attributes of a sample, which can overturn the original decision. These are known as counterfactual explanations. They help users understand ``\emph{why a different prediction was not possible?''} or ``what changes could modify the final prediction?'' Counterfactual explanations can also be illustrated by introducing new example samples that could reverse the decision, known as "what-if" scenarios. Various XAI approaches have been developed to explain "what-if" and counterfactual scenarios ~\cite{van2021interpretable,mothilal2020explaining,wachter2017counterfactual}. Other methods, such as the partial dependence plot (PDP), individual conditional expectation (ICE), and DiCE~\cite{mothilal2020explaining}, are also used for similar purposes. To integrate multiple types of explanations into a unified framework, several open-source toolkits are available, including captum~\cite{kokhlikyan2020captum}, AIX360~\cite{arya2021ai}, and Anchor~\cite{ribeiro2018anchors}.

\subsection{Human-centered XAI}

The progress in the field of XAI, thus far, in generating technical explanations to interpret models for AI practitioners for debugging and improvement, far exceeds the advances made in human-centered perspectives of XAI. However, a significant number of scientific studies have emerged that focus on the user-centric perspective. In this section, we review some notable works on the evaluation of human-centered XAI.

Bell et al.~\cite{bell2022s} demonstrated that the model-agnostic explanations provided by one of the most prominent XAI techniques, SHAP, are not sufficiently comprehensible for general users. They assessed the effectiveness of these explanations through an empirical study involving non-technical participants in two distinct application areas, education, and finance~\cite{bell2022s}. Similarly, Abdul et al.~\cite{abdul2020cogam} explored the trade-off between accuracy and simplicity in explanation presentation and proposed a cognitive generalized additive model (COGAM) for human-centered explanation delivery. A generalized XAI design principle was introduced to facilitate the presentation of local explanations to non-expert users by contextualizing the exploration of feature importance. An empirical study involving more than 80 participants was conducted to evaluate the effectiveness and user satisfaction with the explanations provided.

Chromik et al.~\cite{chromik2021think} investigated the capability of non-expert users to understand and form a mental model of global explanations to comprehend the behavior of a model. Their findings indicated that global explanations are insufficient for understanding the overall model behavior. In a separate study, Hase et al.~\cite{hase2020evaluating} to measured the \emph{simulatability} of various XAI techniques, including LIME, Anchor, prototypes, and decision boundaries, for tabular and textual data. They concluded that LIME achieved good simulatability in a few cases, and prototype models were useful for counterfactual explanations. Another study on simulatability found that current interpretability methods are inadequate in explaining model behavior~\cite{lipton2018mythos,rong2022towards}. However, multiple studies assessing the effectiveness of incorporating XAI approaches in real-world decision-making concluded that the evidence does not align with the goals of XAI~\cite{liao2021human,bansal2021does,poursabzi2021manipulating,wang2022effects}.

\subsection{The Research Gap} 
The related works discussed above, addressing both technical and human-centered XAI, reveal a significant research gap in designing XAI systems for general users. Although current XAI systems can provide interpretable explanations in various forms—including global, local, counterfactual, and example-based explanations—these methods often fail to make predictions and model behavior understandable to non-expert users. Yet, one of the primary goals of XAI is to enable lay users to comprehend the predictions. Furthermore, applications in smart homes, such as energy demand forecasting, consumption practices, smart heating systems for comfortable home environments, and thermal comfort preferences with AI-enabled prediction systems, necessitate diverse perspectives for implementing user-centered XAI adoption.

Consider the complex task of developing a human-centered XAI energy prediction system for a smart home. This task is challenging due to the system's internal complexity and the diverse backgrounds of non-expert users. While significant progress has been made in technical XAI for applications related to classification and regression, the complexity of energy forecasting, which involves two different dimensions—time and characteristics or features—makes it difficult to represent the underlying factors behind the forecast. Moreover, the explanations must be designed and presented in a human-centered manner, which poses additional challenges for creating human-centered explanations in smart home applications.

Another smart home application involves the automatic control of the heating system, which can be based on a prediction model of the inhabitants' thermal preferences. This system generally utilizes physiological and environmental data from the inhabitants, as well as weather data, to predict their thermal comfort preferences~\cite{shajalal2024improved,shajalal2022focus}. Based on these predictions, the smart home application then adjusts the heating system to control the indoor temperature. Since this system utilizes data from various sources, it is crucial for users to understand how and in what context their data are used. Therefore, we require human-centered, easily understandable explanations from such systems to ensure transparency and trust.

\section{Human-Centered Explainability in Smart Home} \label{application-scenario}

Smart homes equipped with AI-driven applications, including energy demand forecasting, consumption routines analysis, heating systems monitoring, and controlling indoor temperature based on occupants' thermal preference prediction systems, are becoming increasingly popular~\cite{shajalal2022towards,jensen2018designing}. These systems often use complex ML-based prediction systems that should be understandable to the household inhabitants~\cite{shajalal2022focus,shajalal2022towards,kim2021explainable}. The adoption of advanced XAI within complex smart home systems, offering human-centered explanations, could be transformative. As a result, non-expert smart home users would gain a clear and concise understanding of how their systems predict consumption and preferences, analyze their data for daily consumption practices, and control the indoor environment to ensure a comfortable living space. Incorporating current XAI techniques and including users in the development loop to consider their perspectives could significantly enhance the provision of human-centered explanations.

\begin{figure*}[!htb]
    \centering \includegraphics[width=0.95\linewidth]{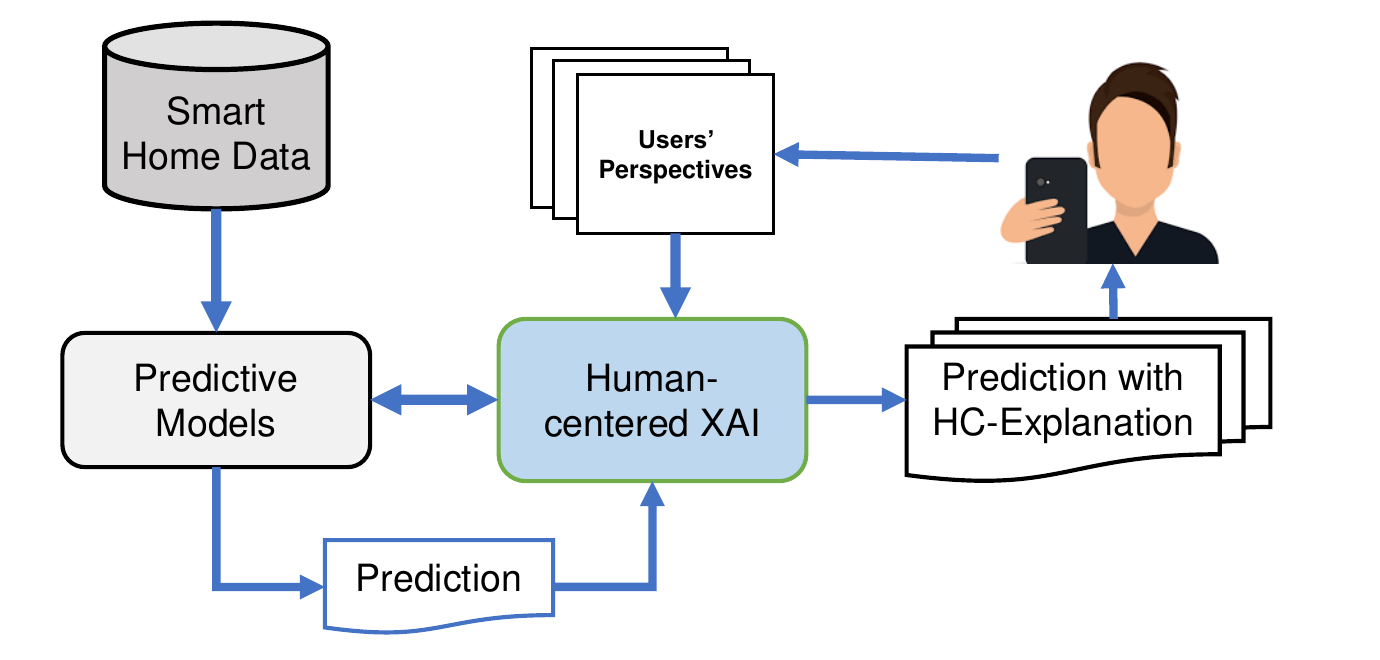}
    \caption{An overview of a human-centered XAI-enabled Smart Home systems}~\label{hcsm}
    \vspace{-1.5em}
\end{figure*}

We present an overview of human-centered XAI in smart home applications with occupants actively involved, as shown in Fig.~\ref{hcsm}. The prediction system should consider the occupants' preferences and intentions to provide meaningful explanations. Initially, ML-enabled predictive models learn from preprocessed data to make future predictions and forecasts. When implementing XAI models to elucidate the decision-making processes of these models and offer explanations for individual predictions, it is crucial to incorporate a human-centered perspective. In this context, human-computer interaction methods are essential for analyzing user feedback. Consequently, effectively presenting these explanations can enhance the adoption of AI-enabled systems in real-world smart home settings. To highlight the importance of human-centered explainability methods in smart home applications, we selected two relevant and extensively investigated problems within the smart home context.

To select the problem domain in the context of smart homes, we focus on applications related to household energy consumption. This emphasis allows us to explore areas that enhance resident comfort while also optimizing energy use. Additionally, smart applications should offer insights to help people optimize and reduce their energy costs. Therefore, we have chosen two applications: household energy demand forecasting and automatic control of heating based on predictions of thermal comfort preferences. Both applications are concerned with energy consumption practices. The first application is an energy demand forecasting system designed to increase awareness and optimize household electricity use. The second involves modeling occupants' thermal comfort preferences to ensure a comfortable and healthy indoor environment. In the remainder of this section, we present two relevant smart home sub-tasks that require human-centered explanations for successful adoption.

\subsection{Household energy demand forecasting}

Energy consumption in residential and commercial buildings significantly exceeds that in other sectors~\cite{shajalal2022focus,shajalal2022towards}. Additionally, the price of energy is continuously increasing worldwide. However, this heightened consumption also results in significant CO2 emissions, posing a serious threat to global warming and the environment~\cite{shajalal2022focus}. Smart home systems address these concerns by providing future energy demand forecasts for households based on historical energy usage data collected from various appliances, thanks to advanced sensor technology~\cite{kim2021explainable}. Such predictions of total energy demand for the upcoming month or week might raise household members' awareness of their energy-related activities, potentially encouraging them to optimize their energy usage.

Nevertheless, energy demand forecasting systems typically rely on highly sophisticated ML and DL techniques~\cite{kim2019electric,kim2021explainable,ehsan2021operationalizing}, which perform complex calculations and lead to opaque decision-making processes. Consequently, some predictions may surprise users with unexpected outcomes. For example, if a forecasting system predicts a high (or low) total energy consumption for washing machines in the next month, users might be taken aback, as they might perceive washing machines to consume less (or more) energy than predicted. In such scenarios, users require easily understandable explanations from the systems to comprehend and build trust in AI-enabled systems, thereby enhancing their adoption.

Unlike other classification and regression tasks, providing explanations for energy demand forecasting systems is non-trivial, as it involves multivariate time series forecasting~\cite{shajalal2022towards}. The explanations in such systems cover two dimensions: they relate to features or attributes and time. Consequently, explainability methods must capture the impact of time and features, making understanding explanations for time series forecasting more challenging for users. Research into human-centered explainability methods in this field is crucial to address this issue. This research will enable inhabitants to understand why a certain energy demand is anticipated for the upcoming month and will facilitate the development of more optimal energy consumption plans based on factual explanations.

\subsection{Occupants' thermal comfort preference modeling}

Indoor thermal comfort is essential for the well-being, comfort, and work productivity of inhabitants~\cite{shajalal2022focus,shajalal2024improved}. With recent advancements in efficient sensors and smart home appliances, AI-driven heating, ventilation, and air conditioning (HVAC) systems can monitor and control the indoor environment. These systems have gained considerable attention for applying machine learning techniques to automatically control parameters related to the comfort of the indoor environment~\cite{shann2017save,shajalal2024improved}. Personal thermal comfort preferences vary widely from person to person, making the prediction of individual preferences crucial for providing occupant-level comfort in households~\cite{shajalal2022focus,shajalal2024improved}. Based on these preference predictions, the heating system can be automatically adjusted to control the temperature at the occupant's location.

Householders often struggle with inconsistent temperatures in their homes, especially during extreme weather conditions, leading to discomfort and high energy costs~\cite{vasseur2019conceptual}. Maintaining a consistent temperature throughout the house can be challenging, causing the HVAC system to work harder to maintain a comfortable temperature using AI-enabled computational models. However, these complex AI systems often lack interpretability. Traditional temperature control systems typically react by adjusting the temperature only after it has already started to fluctuate, resulting in uncomfortable temperature swings and inefficiencies. To address this problem, a more proactive approach to temperature control is needed. Various methods have been introduced to predict personal thermal comfort preferences using complex machine learning and deep learning models~\cite{abdelrahman2022personal,escandon2019thermal,somu2021hybrid,shajalal2022focus}. These automated systems, powered by complex AI models, can monitor and control the indoor environment. However, the decision-making process and the rationale behind specific predictions and actions often remain unclear to the inhabitants, including AI practitioners themselves. As XAI methods progress, it is crucial to make the explanations human-centered, enabling household occupants to understand the reasons behind predictions and the decision-making of the models. This would facilitate the adoption of such complex models and ensure successful smart home applications.

\section{Experiments and Analysis}\label{experiments_xai}
This section presents the details of the experiments we conducted on the two aforementioned smart home scenarios. We carried out experiments by training various predictive models on two distinct datasets collected for both applications. Initially, we trained predictive models and subsequently applied two well-known XAI methods, namely SHAP~\cite{lundberg2017unified} and DeepLIFT~\cite{shrikumar2017learning}. We then analyzed the generated explanations for both smart home applications and sought to identify reasons why these explanations might not be sufficient for smart home users to comprehend. This section presents the experimental results, generated explanations, and their analysis for each scenario.

\subsection{Energy demand forecasting in smart home}

\noindent \textbf{Experimental Settings:} We conducted experiments on the REFIT dataset~\cite{murray2017electrical}, which includes energy consumption data collected from 20 diverse households. The data encompasses various appliances such as the Fridge-Freezer, Tumble Dryer, Washing Machine, Dishwasher, Desktop Computer, Television, Microwave, Kettle, and Toaster, with energy consumption recorded at 8-second intervals. We modeled the weekly energy demand forecasting problem using a classical LSTM-based model. The LSTM-based forecasting model features 10 total features, a sequence length of 7, two hidden layers, 64 hidden units in each layer, 100 epochs, a learning rate of 0.001, and a batch size of 64. \\

\noindent\textbf{Results:} The performance of the LSTM-based forecasting model to predict upcoming weekly energy demand is presented in Table~\ref{REFIT_Performance}. We can see that the performance is quite effective in terms of four different evaluation metrics, including mean squared error~(MSE), Root-MSE~(RMSE), mean average error~(MAE), and Mean absolute percentage error~(MAPE). The performance across different households varied widely. For house 5, the forecasting performance is better than that of another household in terms of MAPE and MSE. On the other hand, for house 13, LSTM achieved the best performance in terms of MAE and RMSE. However, presenting the forecasting performance here makes sense in that the generated explanations for decisions can better capture facts and reasons. 

\begin{table}[!t]
    \centering
    \caption{Prediction performance in forecasting energy demand on 4 different households of REFIT dataset}
    \vspace{0.5em}
    \begin{tabular}{|c|c|c|c|c|}
    \hline    
        \textbf{House} & \textbf{MAE} & \textbf{MAPE} & \textbf{MSE} & \textbf{RMSE} \\ \hline        
        House 2 & 0.1351  & 0.5421  & 0.03  & 0.1752\\ \hline        
        House 5 & 0.0768 &  0.4075 & 0.01  & 0.10\\ \hline 
        House 8 & 0.1934  & 0.4618  & 0.0451  & 0.2095\\ \hline
        House 13 & 0.0435 & 0.8175  & 0.003  & 0.0564\\ \hline
    \end{tabular}
    \label{REFIT_Performance}
\end{table}

\begin{figure*}
    \centering
    \includegraphics[width=0.95\linewidth]{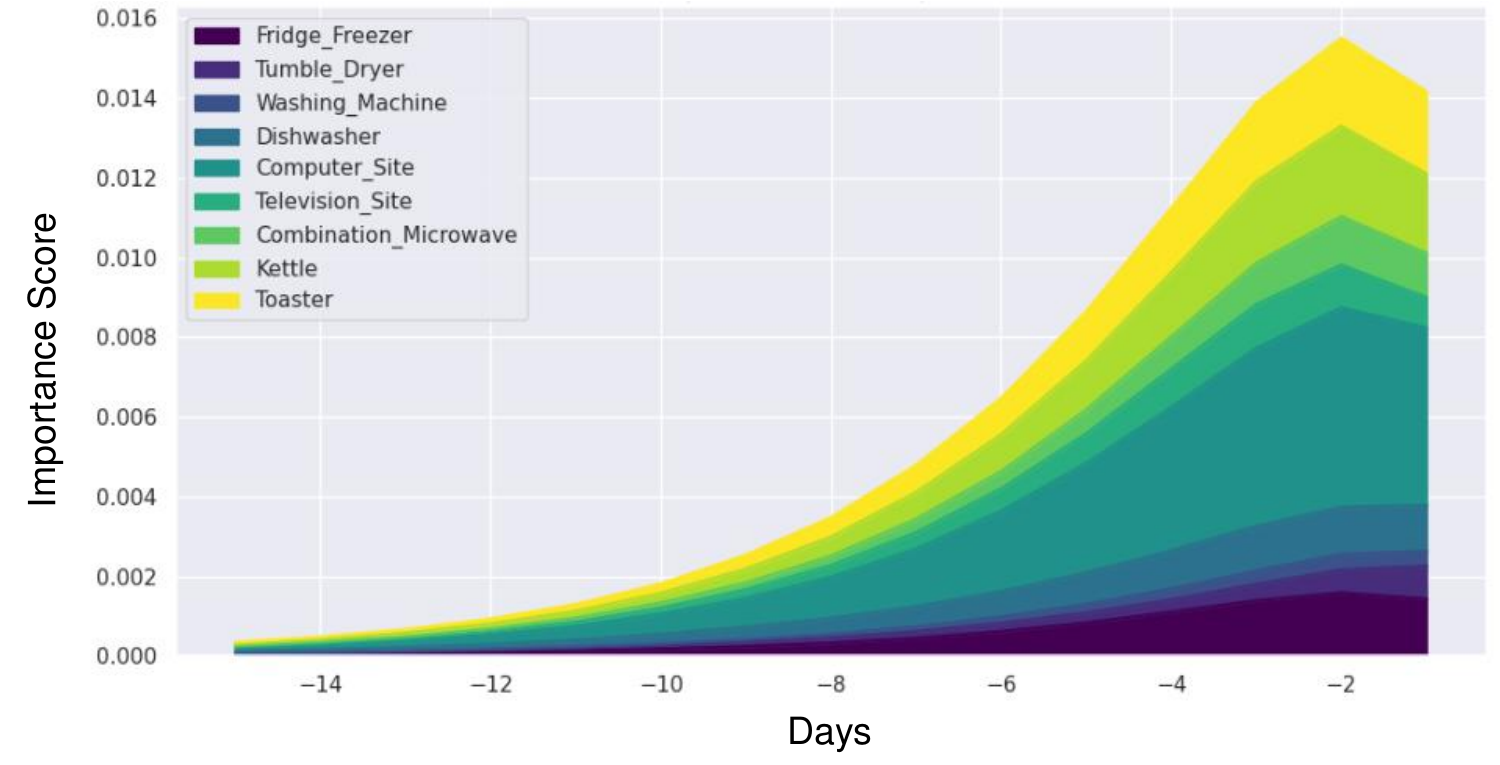}
    \caption{Explanations for weekly energy demand forecasting highlighting the contributions of different appliances.}
    \label{explanation1}
    \vspace{-2.5em}
\end{figure*}

\noindent\textbf{Explainability:} To explain the predictions from the model, we utilized Deep Learning Important Feature~(DeepLIFT), which approximates Shapley values to provide an explanation. This method combines the contributions of different appliances towards the overall prediction, thereby informing users about the activities responsible for the total energy consumption. The explanations for weekly energy forecasting, presented in Figure~\ref{explanation1}, show how the contributions of different appliances vary over time. These explanations are fairly technical, indicating that the contributions of different appliances change with time. Previous studies have shown that general users often struggle to understand even straightforward explanations generated for binary classification tasks.

\begin{figure}
    \centering
    \includegraphics[width=0.95\linewidth]{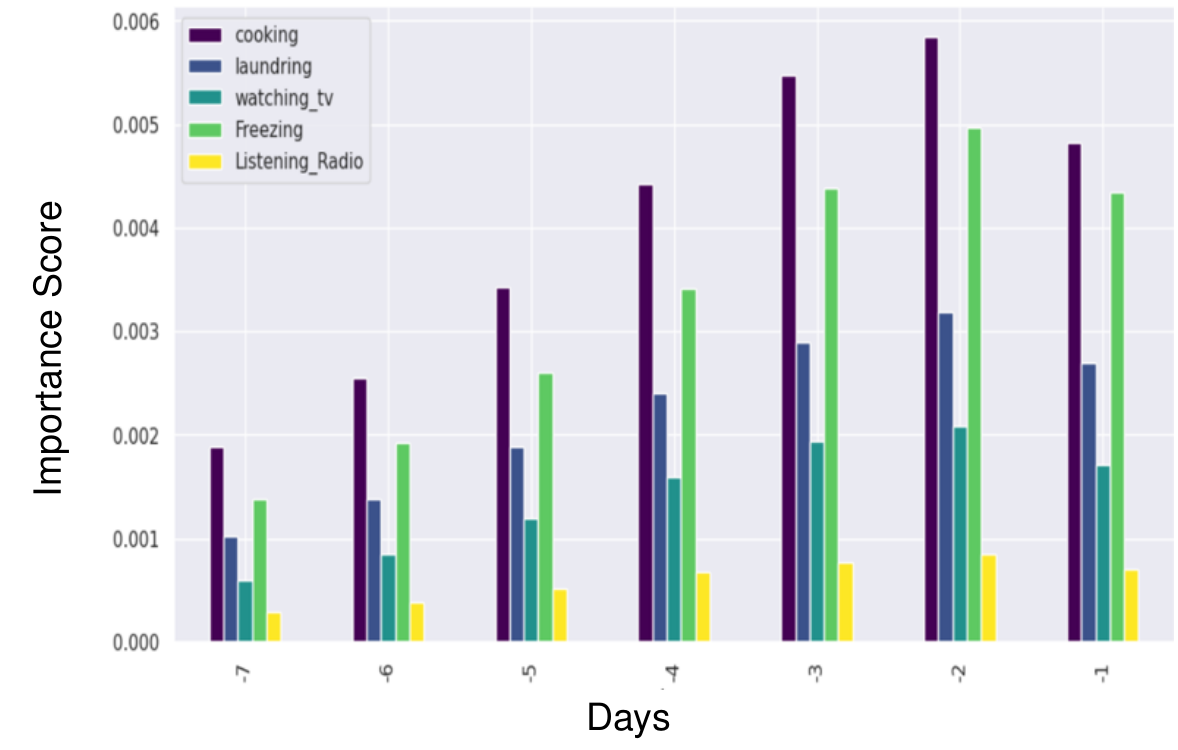}
    \caption{Explanations for weekly energy demand forecasting highlighting consumption activity corresponding to the time (day)}
    \label{explanation2}
    \vspace{-2em}
\end{figure}

On the other hand, energy demand forecasting is an even more challenging task, and the explanations provided differ significantly. The dimension of these explanations also relates to time. As a result, there is a need for explanations that help laypeople understand how these AI-enabled forecasting systems make decisions in their homes. To simplify, we sum up the contributions of different appliances and present the explanation using a bar chart. Figure~\ref{explanation2} shows the contributions of different household activities responsible for overall energy consumption, with cooking activities expected to consume the most energy. What use are these explanations to a user? While they provide some insight into the activities most responsible for energy use, they do not enable users to optimize their energy consumption practices with the current form of explanations. It is also clear that understanding explanations generated by a single successful XAI technique can be challenging. The impact of different activities on overall consumption varies across different days. Non-expert users may develop a negative perception when confronted with such complex explanations. Therefore, explanations should be easy to follow, trustworthy, and tailored to a human-centered perspective to facilitate sense-making.

\begin{table}[!htb]
    \centering
    \caption{The performance of different classical machine learning models on predicting personal thermal comfort preference. The best results are in \textbf{bold}.} 
    \vspace{0.5em}
    \begin{tabular}{|c|c|c|c|c}
    \hline
    \textbf{Model} & \textbf{Kappa} & \textbf{Accuracy} & \textbf{AUC} \\ \hline 
    Decision Tree &0.6609 &0.8315 &0.8810 \\ \hline 
    Support Vector Machine &0.5470 &0.8167 &0.9198 \\ \hline 
    K-nearest Neighbor &0.3810 &0.7589 &0.8101 \\ \hline 
    Gaussian Naive Bayes &0.4311 &0.7148 &0.7689  \\ \hline 
    XGBoost &\textbf{0.6774} &\textbf{0.8457} &\textbf{0.9487} \\ \hline 
    Random Forest &0.5901 &0.8195 &0.9258\\ \hline
    \end{tabular}
    \label{Thermal_Comfort_performance}
\end{table}

\subsection{Personal thermal comfort preference prediction}
\noindent\textbf{Experimental Settings:}
For the second smart home sub-task, "\textit{thermal comfort preference modeling}," we conducted experiments using a wearable dataset collected by UC Berkeley. The dataset originates from a field experiment involving 14 subjects living in Berkeley and San Francisco~\cite{shajalal2024improved}. It contains a total of 3848 samples from all participants, categorized into their thermal comfort preferences: "Cooler" (class 0), "No Change" (class 1), and "Warmer" (class 2). Further details about the dataset can be found in~\cite{shajalal2024improved}. After preprocessing the values of the features, we applied a feature selection technique to identify relevant features, resulting in the selection of 32 different features~\cite{shajalal2024improved}. We then trained six different prominent classical machine learning models, including Decision Tree~(DT), Support Vector Machine~(SVM), K-nearest Neighbor~(KNN), Gaussian Naive Bayes~(GNB), XGBoost~(XGB), and Random Forest~(RF). 

\noindent\textbf{Results:} The performance of all trained machine learning models is presented in Table~\ref{Thermal_Comfort_performance}. We evaluated their performance using metrics such as Cohen's Kappa, Accuracy, and Area Under the Curve (AUC). Given that the dataset was quite imbalanced in terms of the number of samples for different classes, we employed metrics that can effectively evaluate the performance of ML classifiers on imbalanced data. For this purpose, we applied Cohen's Kappa and AUC. From Table~\ref{Thermal_Comfort_performance}, it is evident that XGBoost achieved better performance across all evaluation metrics. The performance of the other five models was also quite consistent compared to the baseline models on the same dataset by~\cite{shajalal2024improved}. \\

\begin{figure}[!t]
    \centering
    \includegraphics[width=0.95\linewidth]{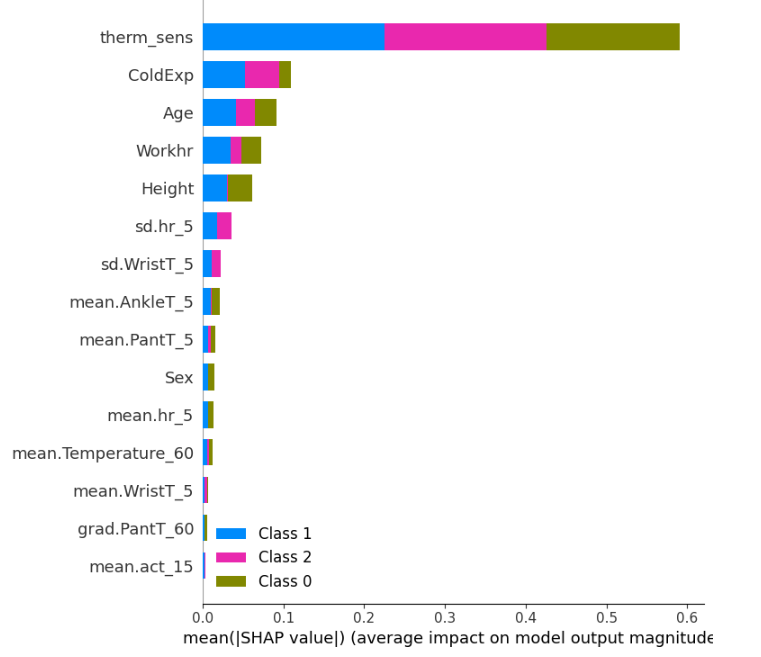}
    \caption{Global explanation for personal thermal comfort preference prediction highlighting model's overall priorities.}
    \label{explanation3}
    \vspace{-1.5em}
\end{figure}

\begin{figure}
    \centering
    \includegraphics[width=0.95\linewidth]{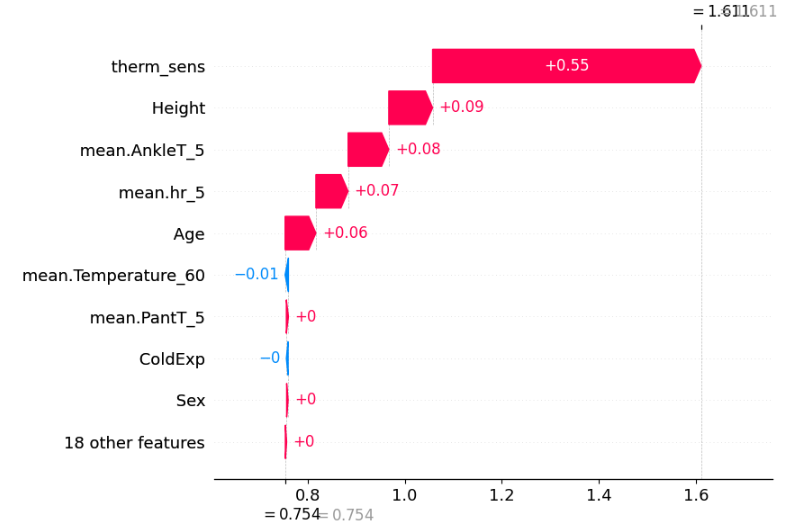}
    \caption{Explanation for a decision that predicts the occupants felling ``\textit{warmer}''.}
    \label{explanation4}
    \vspace{-1.5em}
\end{figure}

\noindent\textbf{Explainability:}
We incorporated the prominent XAI method SHAP to generate explanations. We illustrated the global and local explanations in Figure~\ref{explanation3} and \ref{explanation4}. Figure~\ref{explanation3} shows the global explanations, indicating that thermal sensitivity is the most important feature for modeling personal thermal comfort. The next most important features are cold experience, age, weight, and work hours. As an AI practitioner, one can understand which features the model prioritizes for the overall decision. However, as an inhabitant of the household, some might struggle to derive meaningful insights from this explanation.

In Fig.~\ref{explanation4}, we illustrate the explanations for a particular subject's thermal preference at a certain time in terms of a waterfall plot; these explanations follow the global explanations in a broader sense. We can observe that the most important features are thermal sensitivity, height, and temperature in the ankle. Once again, developers can leverage these insights to enhance the models' performance by canonicalizing and modifying them. Nevertheless, general occupants may find it challenging to derive actionable insights from this information.


\section{Challenges and HCI Techniques for Human-centered Explainability}\label{challenge_direction}
After carefully analyzing the current literature on XAI and human-centered XAI (Section~\ref{HCXAI-SOTA}) and generating explanations using two established XAI methods across two different smart home scenarios, we have identified challenges that need addressing to achieve human-centered XAI in these applications. Smart home applications span various problem domains, including classification (e.g., PTC preference prediction), time-series forecasting (e.g., energy demand forecasting, energy billing), and regression (e.g., activity recognition). Thus, the challenges for achieving user-centric explainability must be tailored to each specific domain. While technical explanations are useful for debugging and optimizing model performance, they do not typically aid general users' decision-making. If explanations do not meet users' needs, this can result in a lack of trust in the model and resistance to its use~\cite{arrieta2020explainable}. Human-centered XAI tools, designed with end-users in mind and providing contextually relevant and understandable explanations, can overcome these challenges~\cite{ehsan2022human}. To offer human-centered explanations for smart home systems, it is crucial to employ user-focused XAI tools, provide understandable explanations, and address the users' inquiries and concerns~\cite{ehsan2022human}. Addressing the following challenges will help in generating human-centered, XAI-enabled explanations, ultimately improving the adoption of AI systems in real-world applications.

\subsection{Challenges in achieving human-centered explainability}

When discussing the challenges of human-centered explainability, we adopt the notions of syntax, semantics, and pragmatics, which are traditionally utilized in linguistics but are effectively applicable in identifying challenges in smart home explanations. These concepts aid in understanding and improving how users design and perceive explanations.

\textbf{The syntax level of explanations} refers to the presentation and visual encoding of explanations, emphasizing user-friendly choices in colors, fonts, layouts, and chart design to minimize cognitive load and simplify interactions with smart devices \cite{Spence_2001}. Although charts generated by XAI frameworks such as SHAP or LIME are highly accurate, they tend to be mathematical and unengaging. In technical domains, this might not pose a problem; however, for lay smart home users, factors such as aesthetics, playfulness, comprehensibility, appropriate measurement units, and careful wording are critical \cite{Castelli_Stevens_Jakobi_2019}. For instance, Schwartz et al. \cite{Schwartz_Stevens_Ramirez_Wulf_2013} demonstrate that technical units like \textit{kWh} (kilowatt-hour) or \textit{kg CO2 eq} (equivalent to the effect of one kg of CO2 emission) are too complex, whereas laypersons typically prefer money as a well-known, easy-to-interpret unit. Another challenge in the visual design of explanations is the small screen size \cite{Spence_2001}, as most people interact with their smart home through smartphones or wall-mounted interfaces, necessitating the simplification of complex explanations for small-screen visualization.

\textbf{The semantic level of explanations} concerns how explanations are interpreted and what mental models are generated \cite{Spence_2001}. In the context of smart homes, for instance, many individuals possess incorrect mental models of heating systems, leading to improper heating behaviors \cite{Kempton_1986}. Therefore, explanations must be mathematically precise and assist users in developing accurate mental models. In this regard, Schwartz et al. \cite{Schwartz_Stevens_Ramirez_Wulf_2013} demonstrate that people often rely on ethno- or folk-methods to construct their mental models. Regarding domestic energy consumption, people interpret the information provided by eco-feedback systems using money as the preferred unit to assess appliance consumption, relate consumption to their habits, or compare their consumption with others' \cite{Schwartz_Stevens_Ramirez_Wulf_2013}. Explanations should leverage these folk methods to help people build accurate mental models \cite{Kempton_1986, Schwartz_Stevens_Ramirez_Wulf_2013}.

\textbf{The pragmatic level of explanations} refers to the context-dependent, practical significance of explanations as they apply to the user’s daily life. For example, explanations can serve various purposes, such as building trust in smart home systems \cite{Jakobi_Stevens_Castelli_Ogonowski_Schaub_Vindice_Randall_Tolmie_Wulf_2018}, increasing the energy literacy of residents \cite{Schwartz_Denef_Stevens_Ramirez_Wulf_2013}, supporting reflection on wasteful consumption habits \cite{Schwartz_Stevens_Ramirez_Wulf_2013}, and prompting actions to detect and replace inefficient appliances. Additionally, they enhance predictability and accountability \cite{Jakobi_Stevens_Castelli_Ogonowski_Schaub_Vindice_Randall_Tolmie_Wulf_2018} and support the co-performance of controlling domestic appliances \cite{Kuijer_Giaccardi_2018}.

Explanations should be tailored to these pragmatic considerations to be effective. For example, simple recommendations accompanied by "what-if" explanations \cite{wachter2017counterfactual} are more effective for actions such as detecting and replacing wasteful appliances. In contrast, more elaborated, cause-effect-oriented explanation approaches \cite{hoffman2017explaining} are more helpful in enhancing energy literacy. By considering all three aspects—syntax, semantics, and pragmatics—we can better identify and overcome the specific challenges associated with explaining smart home technologies, making them more user-friendly and aligned with human-centered design principles.

In the following, we discuss these challenges across three scenarios.

\begin{enumerate}
\item \textbf{Making predictions and autonomous actions interpretable for users:} In the context of energy demand forecasting, XAI tools can provide understandable explanations for the predictions made by the forecasting model, highlighting the most influential factors contributing to the prediction. These explanations should be presented in a user-friendly format that aligns with the user's intuition and understanding of the problem. Visualizations such as graphs and charts can be used to illustrate the data and make it easier for users to comprehend the predictions. In the case of an autonomous action performed by an AI-enabled system (e.g., adjusting the room temperature based on predicted thermal comfort preferences), the system should provide clear explanations for why that particular action was taken.

\item \textbf{Providing insights for an optimal energy consumption plan:} Human-centered, XAI-powered smart home applications should offer users insights into their optimal energy consumption plan by considering their energy usage patterns, environmental conditions, and preferences. The system might involve developing a personalized energy optimization plan, such as adjusting the temperature based on the user's daily routines or turning off lights in unoccupied rooms. The human-centered XAI-enabled system should provide suggestions and explanations for energy-saving practices and offer feedback on the impact of these practices on energy consumption.

\item \textbf{Making users aware of energy consumption:} XAI tools should help users understand how their energy consumption patterns affect the home environment and their energy bills. This can be achieved by providing real-time feedback on energy usage, highlighting areas where energy could be conserved, and suggesting energy-efficient practices. For instance, the XAI tool can send alerts or notifications to remind users to turn off lights or appliances when not in use.
\end{enumerate}

\begin{figure*}[!h]
    \centering \includegraphics[width=0.9\linewidth]{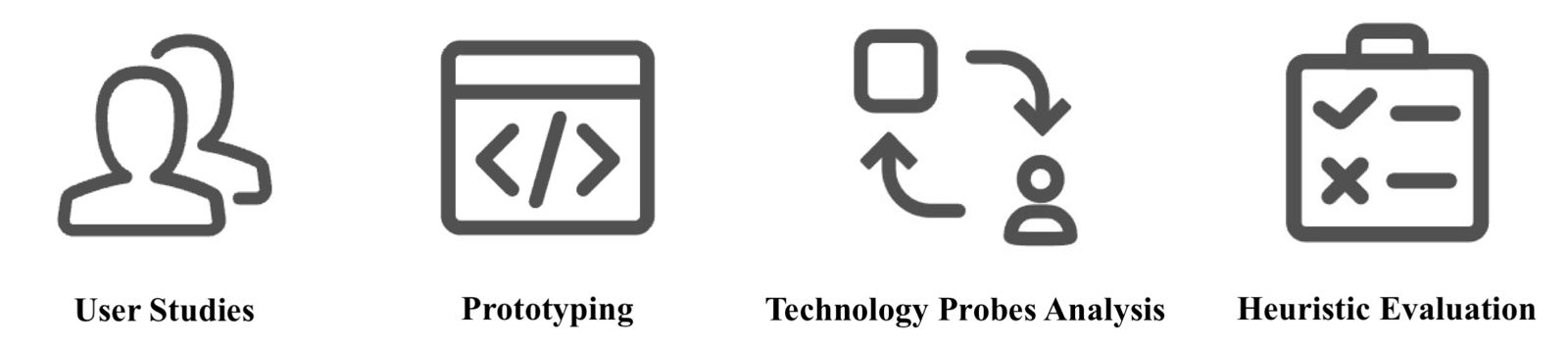}
    \caption{HCI techniques to enhance human-centered explainability}~\label{HCI}
    \vspace{-3em}
    \end{figure*}

\subsection{HCI techniques to enhance human-centered explainability} \label{keypoints}

Focusing on human-centered XAI can align with users' needs and preferences, increasing their trust and understanding of smart home AI systems~\cite{ehsan2022human}. This alignment is particularly significant as we strive for a more energy-efficient and comfortable living environment with smart home systems. Thus, Incorporating HCI methodologies becomes effective in advancing smart home, human-centered XAI implementation. The following non-linear process in Fig.~\ref{HCI} illustrates this approach: User Studies~\cite{rohde2017grounded}, Prototyping~\cite{nissinen2015user}, Technology Probes Analysis~\cite{hutchinson2003technology}, and Heuristic Evaluation~\cite{nielsen1990heuristic}.

\begin{enumerate}
    
\item \textbf{User Studies:}
User studies employ detailed observation, daily tracking, and interviews based on praxeological grounded design methods to understand contextual user practices in smart home environments \cite{rohde2017grounded}. This approach aims to gather deep insights into how users interact with XAI artifacts of smart home technology, facilitating a refined understanding of social practices. By analyzing these interactions, developers can align XAI systems more closely with user habits and expectations. Such alignment increases user trust and enhances system transparency. These insights are crucial for tailoring AI functionalities to manage home heating or cooling systems efficiently. For example, based on collected data, AI could predict and adjust indoor temperatures to optimal levels just before users return home or during specific weather conditions, thus maximizing comfort without the need for manual adjustments \cite{shajalal2022towards, shajalal2024improved, shajalal2022focus}. These adaptive adjustments contribute significantly to improving energy efficiency and overall user comfort.

\item \textbf{Prototyping:} 
Developing several low-fidelity prototypes offers a cost-effective way to explore different designs and continuously gather user feedback. This process is characterized by iterative testing and refinement cycles, heavily using participatory design principles to align with user expectations and improve usability \cite{nissinen2015user}. Through prototyping, designers can quickly adapt and evolve XAI features based on real user feedback, ensuring the system is both intuitive and directly useful to end-users. Such an approach allows developers to fine-tune how XAI communicates energy-saving decisions to users. For example, the explanations might suggest reducing electricity use during off-peak hours as a cost-saving measure. By clearly explaining the rationale behind such recommendations, prototyping enhances the usability of XAI, making it easier for users to trust and follow the AI's guidance. This process not only aids in reducing energy costs but also helps in educating users about efficient energy practices.

\item \textbf{Technology Probes Analysis:}
Technology probes are invaluable for understanding how users interact with and respond to XAI systems within smart homes, especially focusing on aspects such as interpretability, responsibility, and relevance of AI explanations. By deploying technology probes, such as smart meters that track energy usage and provide feedback and advice based on AI analysis, developers can gather rich data on user behavior and preferences \cite{hutchinson2003technology}. These probes reveal user reactions to automated suggestions for optimizing energy consumption. The insights gained from technology probes allow developers to refine human-centered XAI explanations, ensuring they are both meaningful and actionable. This process enhances the system's usability and boosts users’ understanding of and trust in the AI explanations, leading to more effective and sustainable smart energy management.

\item \textbf{Heuristic Evaluation:}
Heuristic evaluation involves collaboration with subject matter experts, including HVAC, electrical, and social engineers, to assess AI systems' responsibility and user-centric design. This evaluation focuses on key aspects such as transparency, user control, and ethical considerations. Experts examine whether an AI system’s explanations for recommending specific energy-saving measures are understandable and align with user values \cite{nielsen1990heuristic}. Such evaluations are crucial because they help ensure that AI systems are designed with a strong emphasis on user-centric principles, which promotes a better understanding of and trust in the technology. Heuristic evaluation is particularly effective at identifying aspects of AI explanations that may not be evident through end-user testing alone. Doing so aids users in making informed decisions about their energy use, like understanding why certain settings are recommended for maximizing thermal comfort without excessive energy use \cite{shajalal2022towards, shajalal2024improved, shajalal2022focus}.

\end{enumerate}

\section{Conclusions and Future Directions} \label{conclusion}
Our research strives to achieve human-centered XAI for smart home applications, aiming to make complex AI-driven models understandable to laypeople. Through experiments on two smart home sub-tasks, we demonstrated the challenges associated with understanding decisions from such applications. We argue that current explanation generation techniques are insufficient for making general users comprehend these decisions. By identifying major challenges, we highlight the need for human-centered explainability and discuss how these challenges can be addressed using various human-computer interaction (HCI) techniques, including user feedback, co-performance considerations, and expert-user co-design. Future human-centered XAI can apply HCI techniques to understand users' requirements and preferences better, enabling them to understand decisions from AI-driven systems. Based on the outcomes of these techniques, we aim to develop human-centered explanations that facilitate user understanding and action. 

However, challenges persist in generating human-centered explanations that foster trust and interpretability in AI systems. Future research directions include:
\begin{itemize}
\item Exploring natural language explanations and interactive interfaces
\item Developing standardized frameworks to evaluate the general user experience of human-centered explainability
\item Integrating HCI methodologies into the XAI development life-cycle to ensure user-centered and effective explanations
\end{itemize}

Future research on emerging user interfaces, including ubiquitous and pervasive technologies, can also advance the current state of the art on human-centered XAI in the context of the smart home energy domain. As we envision the integration of smart devices such as voice assistants, smart watches, and embedded sensors into everyday environments, these technologies provide a rich platform for deploying human-centered solutions to benefit sustainability. Such interfaces can offer intuitive and context-aware interactions, making AI explanations part of the natural user environment and activities.

\section*{Acknowledgment}
This project has received funding from the European Union's Horizon 2020 research and innovation programme under the Marie Skłodowska-Curie grant agreement No 955422.

\bibliographystyle{unsrt}
\bibliography{reference}
\end{document}